\documentclass[aps,onecolumn,showpacs]{revtex4}
\usepackage{dcolumn}
\usepackage{graphicx}
\usepackage{amsmath}
\usepackage{amsfonts}
\usepackage{amssymb}
\usepackage{psfrag}
\usepackage{wrapfig}
\usepackage{subfigure}
\usepackage{makeidx}
\usepackage{bm}
\usepackage{epsf}

\begin{document}

\title{Novel Localized Waves in a Two-mode Nonlinear Fiber with High-order Effects}
\author{Li-Chen Zhao$^{1}$, Zhan-Ying Yang$^{1}$}
\author{Liming Ling$^{2}$}\email{lingliming@qq.com}
\address{$^1$Department of Physics, Northwest University, Xi'an
710069, China}
\address{$^2$Department of Mathematics, South China University of Technology, Guangzhou 510640, China}

 %%%%%%%%%%%%%%%%%%%%%%%%%%%%%%%%%%%%%%%%%%%%%%%%%
\date{July 29, 2013}
\begin{abstract}

 We study on rational solutions on nonzero background of coupled Sasa-Satsuma
 equations through Darboux transformation method, which take into account
 third order dispersion, the term with
self-frequency shift, and the term describing self-steepening
corrections to the cubic nonlinearity. We find there are
 some new types of localized waves in the coupled system, such as
 dark-antidark soliton pair, W-shaped soliton,
 dark W-shaped soliton, and the combined localized waves of them.
 The results indicate that there are abundant novel localized
 waves in the two-mode fiber with these high-order effects, which would inspire
 experimental realization in the related physical systems.

\end{abstract}

 \maketitle

\section{Introduction}
Nonlinear localized waves are arising from the interplay between
self-focusing (self-defocusing) and dispersion effect, and could be
one of the intense studies in nonlinear science. The well-known ones
of the scalar nonlinear Schr\"odinger(NLS) equation  have been
studied in many different systems including plasmas, optical fibers
and cold atoms, mainly including bright solitons \cite{1,2,3,4},
dark solitons \cite{Burger,6,7,8,9,Wu}, Akhmediev breather (AB)
\cite{N.Akhmediev}, and rogue waves (RW) \cite{Shrira} .  Moreover,
the AB and RW have been observed in one-mode nonlinear
fibers\cite{Kibler} and in water wave tank\cite{A. Chabchoub}
experimentally. For nonlinear fiber, the simplified NLS just contain
group velocity dispersion (GVD) and its counterpart self-phase
modulation (SPM). But for ultrashort pulses, in addition to the SPM,
the nonlinear susceptibility will produce higher order nonlinear
effects like the self-steepening (otherwise called the Kerr
dispersion) and the stimulated Raman scattering (SRS). Apart from
GVD, the ultrashort pulse will also suffer from third order
dispersion (TOD). These are the most general terms that have to be
taken into account when extending the applicability of the NLS
\cite{K. Porsezian,Kodama}. With these effects, the corresponding
integrable equation was derived as Sasa-Satsuma(S-S) equation
\cite{SS}. The localized waves of S-S equation have distinctive
properties from the ones of the well-known NLS equation\cite{rws}.

 Recent studies on localized waves have
been extended to multi-component coupled
systems\cite{Kockaert,Kanna,Lakshmanan,Zhao,Becker,Park,Forest},
since a variety of complex systems, such as Bose-Einstein
condensates, nonlinear optical fibers, etc., usually involve more
than one component \cite{Baronio}. It found that there are many new
localized waves in coupled systems which are different from the ones
in scalar systems, such as dark RW\cite{Bludov2,Zhao2}, four-petaled
flower structure RW \cite{Zhao3}, and multi-RW \cite{Zhao3,
Degasperis}. Moreover, different kinds of localized waves can
coexist and interplay with each other in the coupled system, such as
RW interplay with dark soliton, breather, etc.
\cite{Baronio,Ling2,Zhao2,Degasperis2}. These studies indicate that
nonlinear waves in coupled system are much more diverse than the
ones in uncoupled systems \cite{Zhao3,Kevrekidis}.

 In this
paper, we study on analytical localized wave solutions which contain
rational forms of two-component coupled S-S equations through
Darboux transformation method, which can be used to describe the
evolution of optical fields in a two-mode nonlinear fiber with TOD,
SRS, cross-phase modulation, and self-steepening effects. We find
there are
 some new types of localized wave solutions in the coupled system, including
 dark-antidark(D-AD) soliton pair solution, semi-rational, and rational solutions. For
 the D-AD soliton pair solution, there is a dark and antidark soliton pair in each
 component, and the hump or valley in one component corresponds to
 the valley or hump in the other component. For the semi-rational
 solution, we observe one D-AD pair splits into one new D-AD and W-shaped soliton.
 For rational solution, we observe a W-shaped soliton, which is quite
 different from the rational solutions obtained in coupled NLS equations. The W-shaped soliton
  has the identical shape with the well-known NLS RW with maximum peak. Furthermore, we
 find that there are many combined localized waves which are consist of
 W-shaped soltion and dark W-shaped soltion.

\section{The coupled S-S model}
According to the original work of Sasa and Satsuma \cite{SS}, and
the studies on coupled S-S equations \cite{K. Nakkeeran},  the
evolution equations for the optical fields in a two-mode fiber with
the high-order effects mentioned above can be written as

\begin{eqnarray}
&&iE_{1z} + \frac{1}{2} E_{1tt}+(|E_1|^2+|E_2|^2) E_1 +i \epsilon
[E_{1ttt} \nonumber\\&&+ 6 (|E_1|^2+|E_2|^2) E_{1t}+ 3 E_1
(|E_1|^2+|E_2|^2)_t]=0,\nonumber\\
 && iE_{2z} + \frac{1}{2}
E_{2tt}+(|E_1|^2+|E_2|^2) E_2 +i \epsilon [E_{2ttt} \nonumber\\&&+ 6
(|E_1|^2+|E_2|^2) E_{2t}+ 3 E_2 (|E_1|^2+|E_2|^2)_t]=0.
\end{eqnarray}
Here, an arbitrary real parameter $\epsilon $ scales the integrable
perturbations of the NLS equation. When $ \epsilon= 0$, Eq. (1) and
(2) reduces to the standard coupled NLS equations which have only
the terms describing lowest order dispersion and self-phase
modulation. The vector soliton solutions have been presented from
trivial zero solution through performing the Darboux transformation
method in \cite{K. Nakkeeran}. Here, we study localized waves
solution which contains rational solution from plane seed solutions
which can be seen as the backgrounds where nonlinear localized waves
emerge. For coupled S-S equations, the relative frequency has the
real effects for the localized wave dynamics. Therefore, we study on
localized waves on the following backgrounds

\begin{eqnarray}\label{simptrans}
E_{10}(t,z)&=&c_1 \exp{[i \theta_1]}\exp{[\frac{i}{6\epsilon}(t-\frac{z}{18\epsilon})]},\nonumber\\
E_{20}(t,z)&=&c_2 \exp{[i
\theta_2]}\exp{[\frac{i}{6\epsilon}(t-\frac{z}{18\epsilon})]},
\end{eqnarray}
where
\begin{eqnarray}
\theta_1&=& k_1 T+\epsilon k_1^3 z-6 \epsilon k_1
(c_1^2+c_2^2)z,\nonumber\\
 \theta_2&=&k_2 T+\epsilon k_2^3 z-6
\epsilon k_2 (c_1^2+c_2^2)z,\nonumber
\end{eqnarray}
and $T=t-\frac{z}{12 \epsilon}$. $c_1$ and $c_2$ denote the
background amplitude of the two components respectively. $k_1$ and
$k_2$ are the frequencies of the two components respectively.

For coupled S-S equation, there are two component
fields in the system. Therefore, the relative frequency has the real
effects for the localized wave dynamics. We will present the method
to derive more generalized solutions as follows. The Eq.(1) can be simplified as follows through the
proper variable transformation \cite{K. Nakkeeran}
\begin{eqnarray}
&&q_{1Z} + \epsilon [q_{1TTT}+ 6 (|q_1|^2+|q_2|^2) q_{1T}+ 3 q_1
(|q_1|^2+|q_2|^2)_T]=0,\nonumber\\
&& q_{2Z} + \epsilon [q_{2TTT}+ 6 (|q_1|^2+|q_2|^2) q_{2T}+ 3 q_2 (|q_1|^2+|q_2|^2)_T]=0.
\end{eqnarray}
The variable transformation is given as
\begin{eqnarray}
q_1(T,Z)&=&E_1(t,z)\exp{[\frac{-i}{6\epsilon}(t-\frac{z}{18\epsilon})]},\nonumber\\
q_2(T,Z)&=&E_2(t,z)\exp{[\frac{-i}{6\epsilon}(t-\frac{z}{18\epsilon})]},
\end{eqnarray}
where $T=t-\frac{z}{12 \epsilon}$ and $Z=z$. The corresponding
Lax-pair of Eq. (3) can be given as
\begin{eqnarray}
\Psi_T&=&U \Psi, \  \
 \Psi_Z=V \Psi,
\end{eqnarray}
where $\Psi=(\Psi_1, \Psi_2, \Psi_3, \Psi_4, \Psi_5)^T$, and

\begin{eqnarray*}
 U&=&-i\lambda \sigma_3+Q,\\
V&=&\frac{8i}{5}\epsilon \lambda^3\sigma_0+4\epsilon
\lambda^2Q+2{\rm i}\lambda
\sigma_3(Q_T-Q^2)\\&&-\epsilon(Q_{TT}+QQ_T-Q_TQ-2Q^3),
\end{eqnarray*}
where $\lambda$ is the spectral parameter, and
\begin{equation*}
   \begin{split}
    Q&=\begin{pmatrix}
 0 & \mathbf{q} \\
  -\mathbf{q}^{\dag} &0_{4\times 4} \\
\end{pmatrix},\quad \sigma_3=\begin{pmatrix}
                               1 & 0 \\
                               0 & -I_{4\times 4} \\
                             \end{pmatrix},\\ \sigma_0&=\begin{pmatrix}
                               -4 & 0 \\
                               0 & I_{4\times 4} \\
                             \end{pmatrix},\quad
                             \mathbf{q}=(q_1,q_1^*,q_2,q_2^*).
   \end{split}
\end{equation*}

 To derive more generalized solution, we
solve the Lax-pair from more generalized seed solutions. The seed
solutions of Eq. (3) can be written as follows from the seed solution Eq.(2),
\begin{eqnarray}
q_{10}(T,Z)&=&c_1 \exp{[i \theta_1]},\nonumber\\
q_{20}(T,Z)&=&c_2 \exp{[i \theta_2]},
\end{eqnarray}
where 
\begin{eqnarray}
\theta_1&=& k_1 T+\epsilon k_1^3 Z-6 \epsilon k_1
(c_1^2+c_2^2)Z,\nonumber\\
 \theta_2&=&k_2 T+\epsilon k_2^3 Z-6
\epsilon k_2 (c_1^2+c_2^2)Z.\nonumber
\end{eqnarray}
Solving the Lax-pair (5) from the seed solution,
we can derive the localized wave solutions on
arbitrary plane background through the following Darboux
transformation
\begin{eqnarray}
q_{1}&=&q_{10}-\frac{2 i(\lambda-\lambda^{\ast}) \Psi_1\Psi_2^{\ast}} {\sum_{i=1}^5|\Psi_i|^2},\nonumber\\
q_{2}&=&q_{20}-\frac{2 i(\lambda-\lambda^{\ast})
\Psi_1\Psi_4^{\ast}} {\sum_{i=1}^5|\Psi_i|^2}.
\end{eqnarray}
One can derive the localized wave solutions of
Eq.(1) directly though the variable transformation Eq.(4).
 We find there are two kinds of new rational solutions in
the coupled system with some certain conditions on the spectral
parameter or backgrounds' amplitudes and frequencies.

\section{Two kinds of new rational solutions in coupled S-S model}

\subsection{Dark-antidark soliton pair and W-shaped soliton}

For the simplest case $k_1=k_2=0$ which corresponds to that the
frequencies of the two modes are equal, we obtain an exact solution
which contains rational functions as follows
\begin{eqnarray}
E_{1}[t,z]&=&(c_1 + \frac{H_1[t,z]}{G[t,z]})\cdot \exp{[\frac{i}{6 \epsilon}(t-\frac{z}{18\epsilon})]},\\
E_{2}[t,z]&=&(c_2 + \frac{H_2[t,z]}{G[t,z]})\cdot
\exp{[\frac{i}{6\epsilon}(t-\frac{z}{18\epsilon})]}.
\end{eqnarray}
where
\begin{widetext}
\begin{eqnarray*}
H_{1}&=&4 \sqrt{2}c_1 c e^{\sqrt{2} c (t-\frac{z}{12\epsilon})}
\left(-A_1 c_1^2 e^{\sqrt{2} c (t-\frac{z}{12\epsilon})}+A_2 c_1^2
e^{\sqrt{2} c (t-\frac{z}{12\epsilon})}
[-(t-\frac{z}{12\epsilon})+12
c^{2} z \epsilon ]+A c_2^2 e^{8 \sqrt{2} c^3 z \epsilon }\right)\nonumber\\
&& \left(-\sqrt{2} A_1 c+A_2 (-1-\sqrt{2} c (t-\frac{z}{12\epsilon})
+12 \sqrt{2}c_1^2 c z \epsilon +12 \sqrt{2} c_2^2 c
z \epsilon)\right),\nonumber\\
H_{2}&=&4 \sqrt{2} c_1^2 c_2 c e^{\sqrt{2} c
(t-\frac{z}{12\epsilon})} \left(-\sqrt{2} A_1 c-A_2-\sqrt{2}A_2 c
(t-\frac{z}{12\epsilon})+12 \sqrt{2}A_2 c_1^2 c z \epsilon +12
\sqrt{2} A_2c_2^2 c z \epsilon \right)\nonumber\\
&&  \left(-A_1 e^{\sqrt{2} c (t-\frac{z}{12\epsilon})}-A e^{8
\sqrt{2} c^3 z \epsilon } -A_2 e^{\sqrt{2} c
(t-\frac{z}{12\epsilon})}[t-\frac{z}{12\epsilon}-12 c^{2} z \epsilon
]\right),\nonumber\\
 G&=& -2 c^{2} \left(2 A_1^2 c_1^2 e^{2 \sqrt{2}
c (t-\frac{z}{12\epsilon})}+A^2 c_2^2 e^{16 \sqrt{2} c^3 z \epsilon
}\right)+2 A_1 A_2 c_1^2 e^{2 \sqrt{2} c (t-\frac{z}{12\epsilon})}
\left(-\sqrt{2} c-4 c_2^2 (t-\frac{z}{12\epsilon})\right. \nonumber\\
&&\left.+48 c_1^4 z \epsilon +48 c_2^4 z \epsilon -4 c_1^2
(t-\frac{z}{12\epsilon}-24 c_2^2 z \epsilon)\right)-A_2^2 c_1^2 e^{2
\sqrt{2} c (t-\frac{z}{12\epsilon})} \left(1+4
c^{2} (t-\frac{z}{12\epsilon})^2\right. \nonumber\\
&&\left.-24 \sqrt{2} c_2^2 c z \epsilon +576 c_1^6 z^2 \epsilon
^2+1728 c_1^4 c_2^2 z^2 \epsilon ^2 +2 (t-\frac{z}{12\epsilon})
(\sqrt{2} c-48 c_1^4 z \epsilon
-96 c_1^2 c_2^2 z \epsilon -48 c_2^4 z \epsilon )\right. \nonumber\\
&&\left.+576 c_2^6 z^2 \epsilon ^2+24 c_1^2 z \epsilon (-\sqrt{2}
c+72 c_2^4 z \epsilon )\right),\nonumber
\end{eqnarray*}
\end{widetext}
 where $c=\sqrt{c_1^2+c_2^2}.$
 The parameters $A_1$, $A_2$
and $A$ are arbitrary real numbers. We find there are mainly three
different kinds of localized waves for the above solution in
different parameters regimes. The solutions have been verified by
Mathematica software package.

 \emph{Dark-antidark soliton pair solution}---When $A_1\neq0$, $A_2=0$,
 and $A\neq0$, the solution corresponds to dark-antidark soliton
 pair solution. To make the solution more concise, we set $A_1=1, A_2=0,
 A=1$, $c_1=c_2=1$, and $\epsilon=\frac{1}{12}$. The simplified solution
can be given as
\begin{eqnarray}
E_{11}&=&\frac{-6 e^{4 t}+(e^{14 z/3}+2 e^{2 t})^2}{2 e^{4 t}+e^{28 z/3}}\cdot \exp{[2i(t-\frac{2z}{3})]},\\
E_{21}&=&\frac{-6 e^{4 t}+(e^{14 z/3}-2 e^{2 t})^2}{2 e^{4 t}+e^{28
z/3}} \cdot \exp{[2i(t-\frac{2z}{3})]}.
\end{eqnarray}
The evolution of them are shown in Fig. 1. We can see that there are
one valley and one hump on the nonzero background. They are close to
each other and form a stable pair, shown in Fig. 1(a) and (b). In
fact, the similar structure can exist in coupled NLS equations, but
the structure is breathing \cite{Zhao2}. Namely, the valley and hump
in each component are switched with propagation distance for coupled
NLS equations. In contrast, for coupled S-S model, the valley and
hump in Fig. 1 evolve stable and can be seen as dark and anti-dark
soliton (a hump on top of a nonzero flat background) separately.
Namely, we can call them dark-antidark(D-AD) soliton pair solution.
Moreover, the hump or valley in $E_1$ component corresponds to
valley or hump in $E_2$ component, shown in Fig. 1(c). The whole
density distribution can be seen as an anti-dark soliton on nonzero
background (the green solid line in Fig. 1(c)).

\begin{figure}[htb]
\centering
\subfigure[]{\includegraphics[height=56mm,width=75mm]{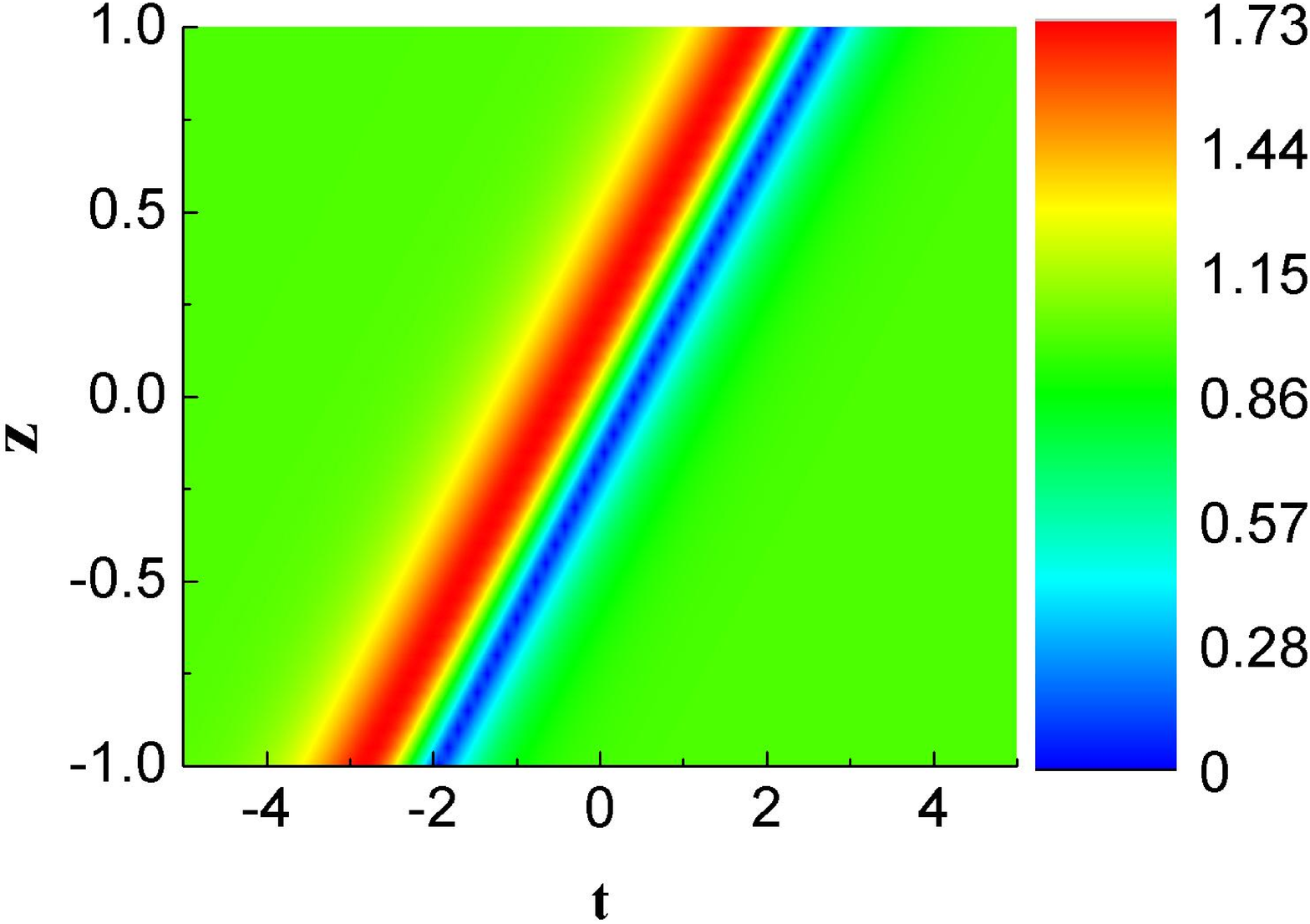}}
\hfil
\subfigure[]{\includegraphics[height=56mm,width=75mm]{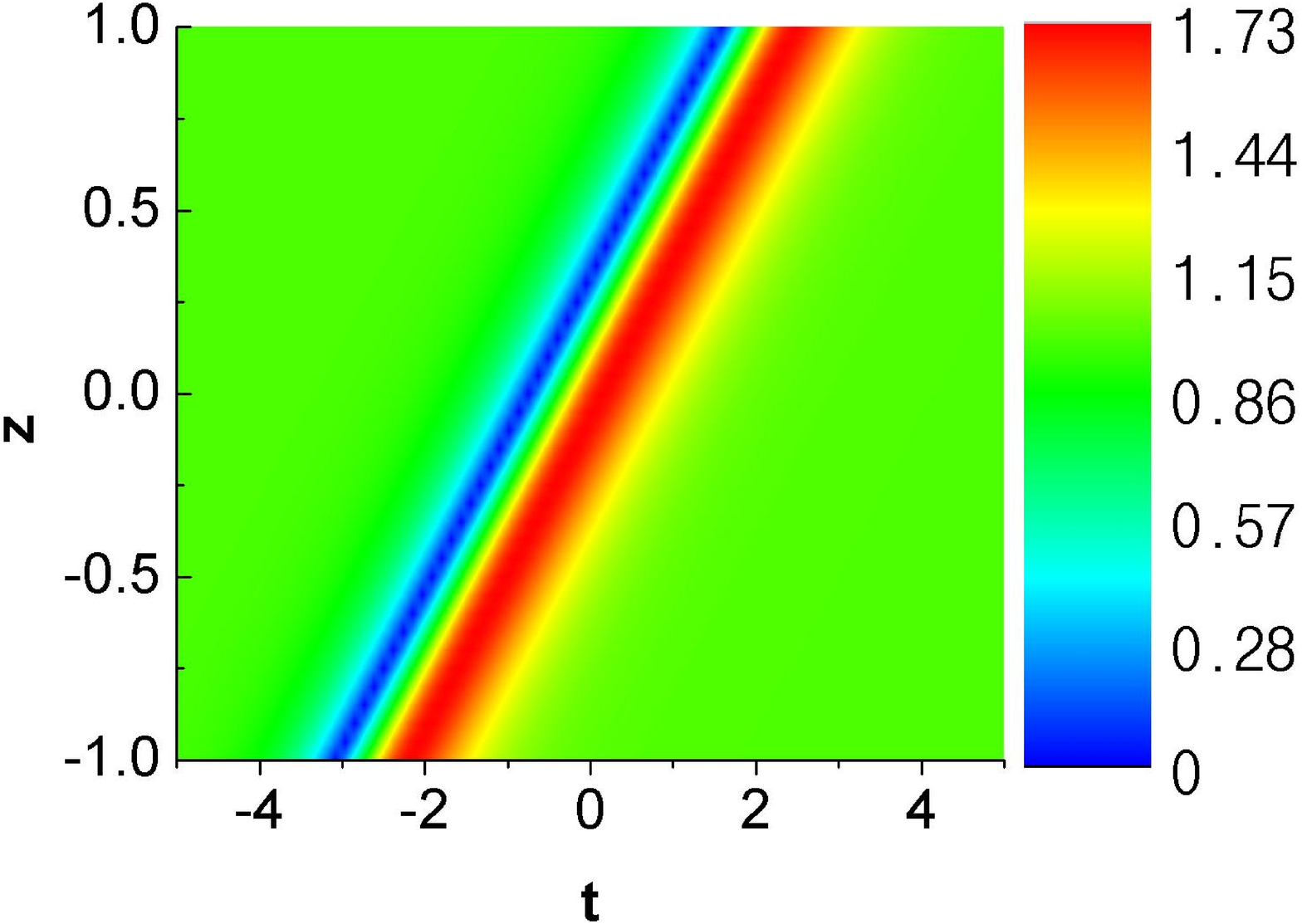}}
\hfil
\subfigure[]{\includegraphics[height=60mm,width=80mm]{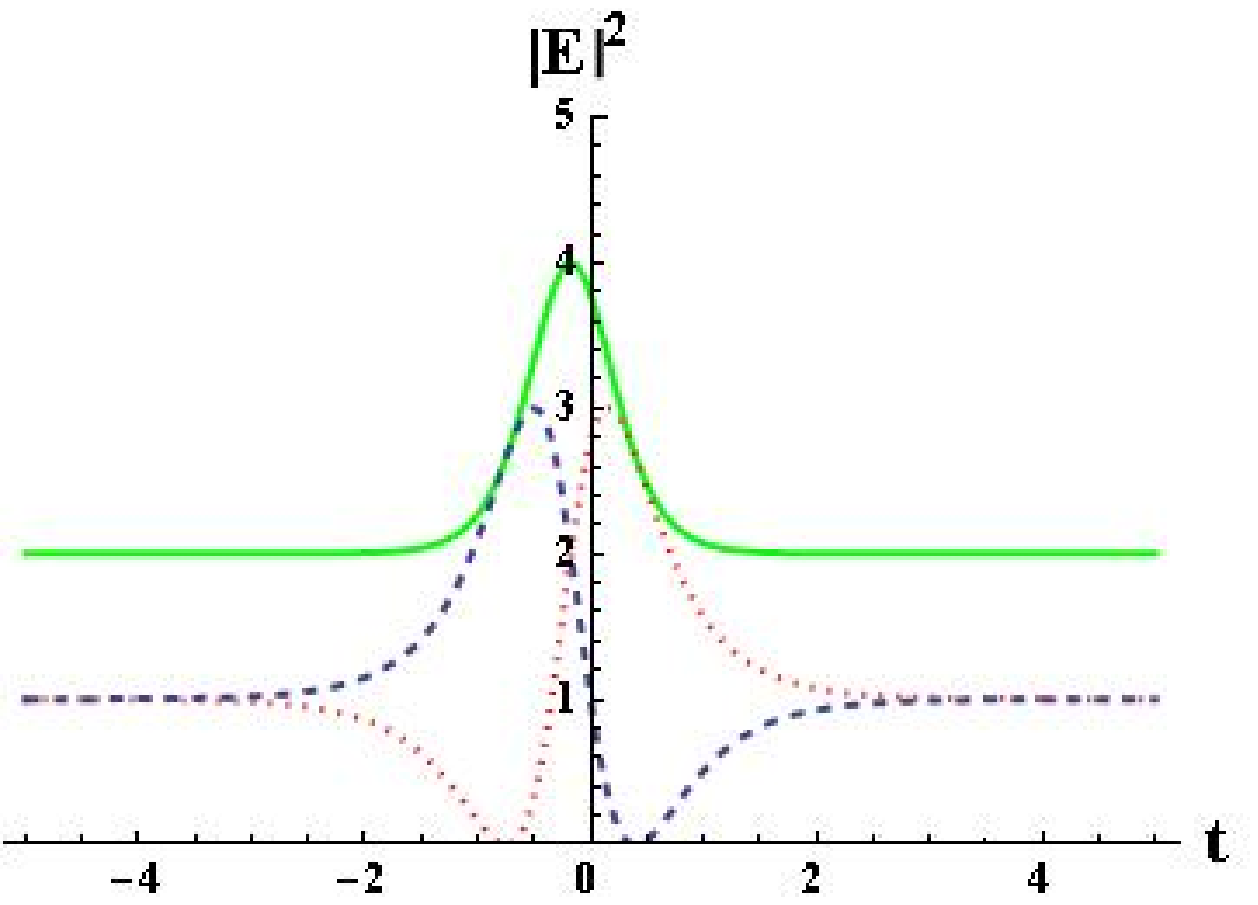}}
\caption{(color online) (a) The evolution of D-AD soliton pair in
component $E_1$. (b) The evolution of D-AD soliton pair in component
$E_2$. (c) The cut plot of the D-AD soliton, the dashed blue line
for $E_1$, the dotted red line for $E_2$, and the solid green line
for the whole density distribution. }
\end{figure}

\begin{figure}[htb]
\centering
\subfigure[]{\includegraphics[height=56mm,width=75mm]{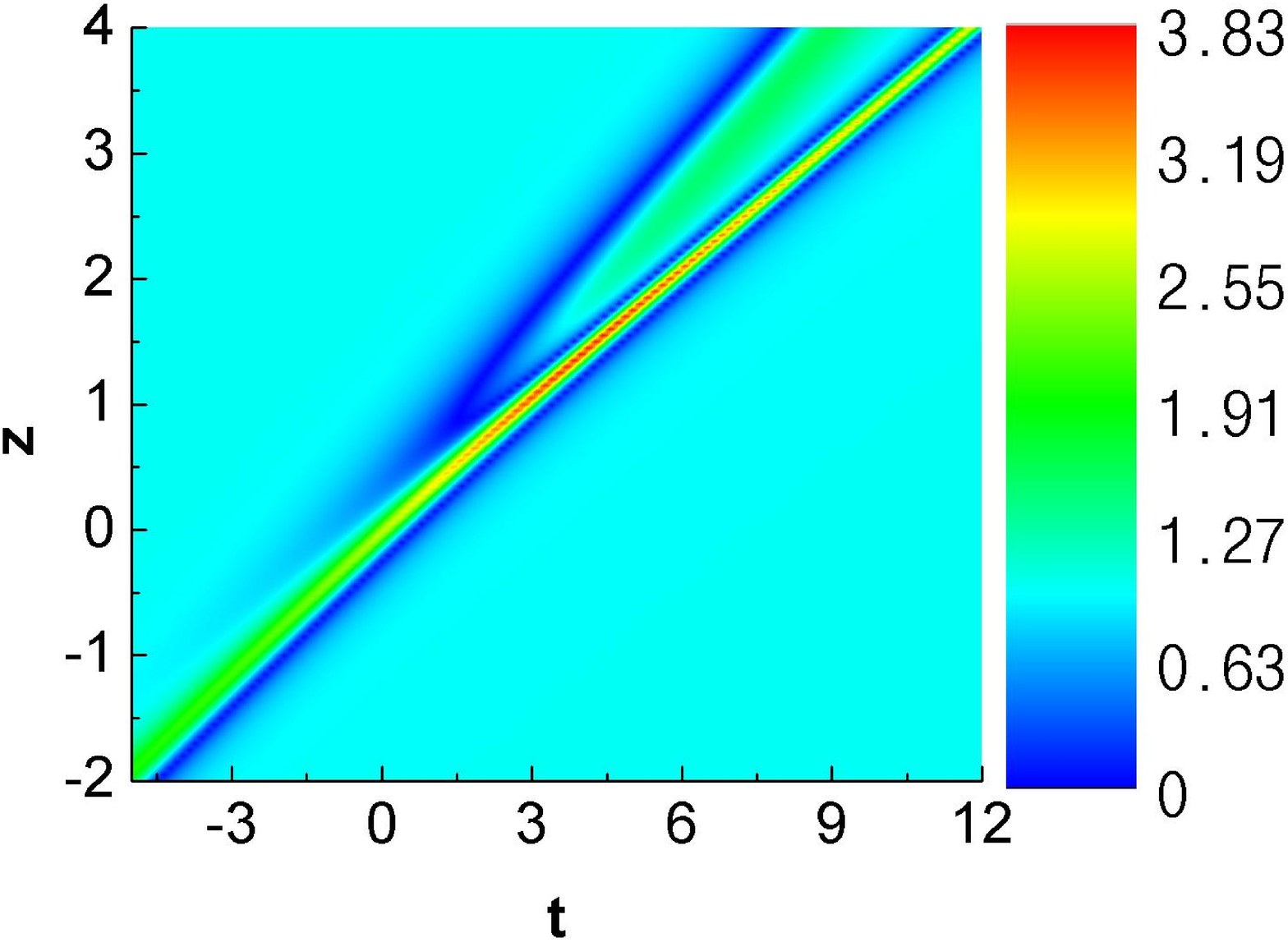}}
\hfil
\subfigure[]{\includegraphics[height=56mm,width=75mm]{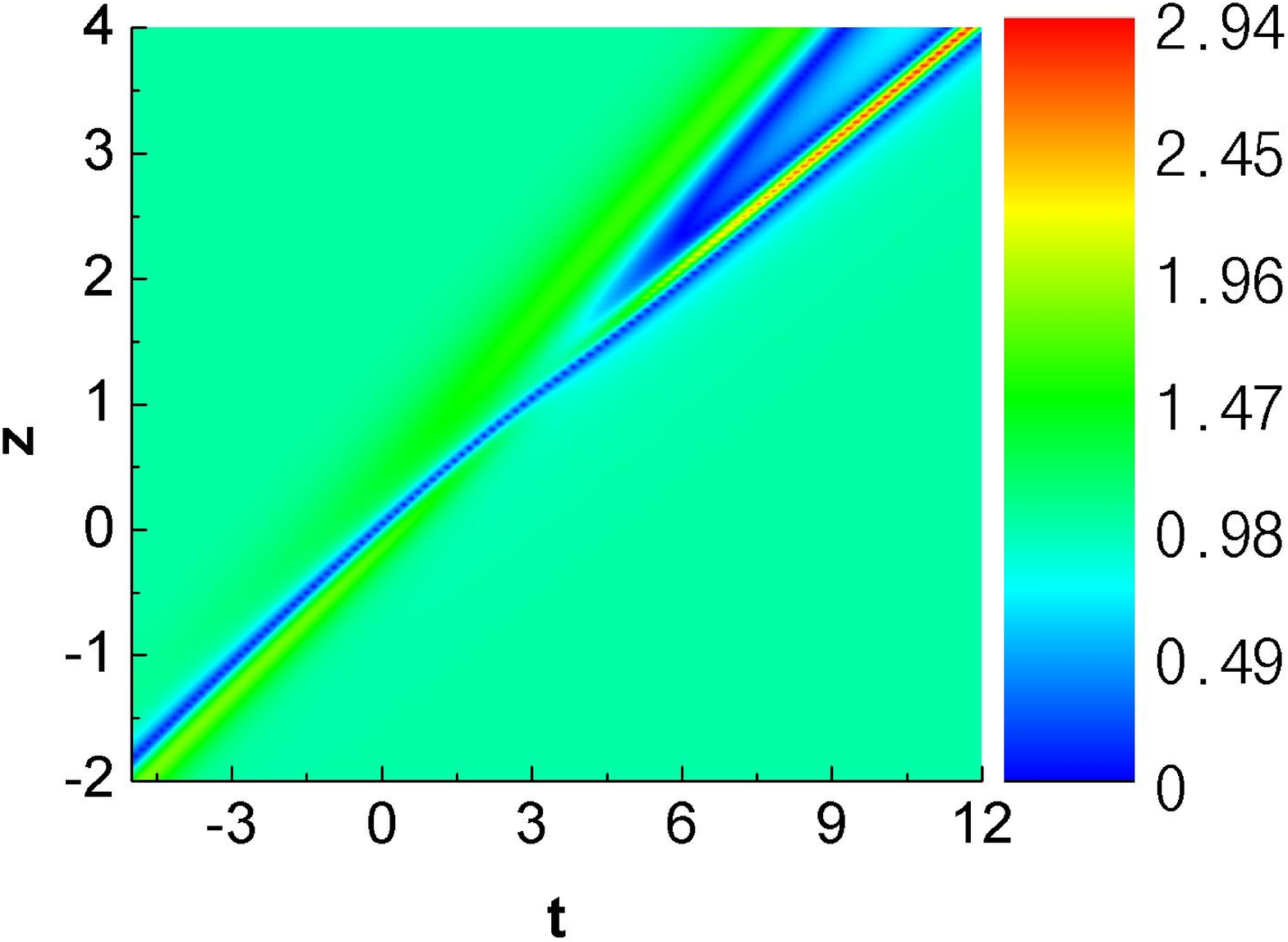}}
\hfil
\subfigure[]{\includegraphics[height=52mm,width=70mm]{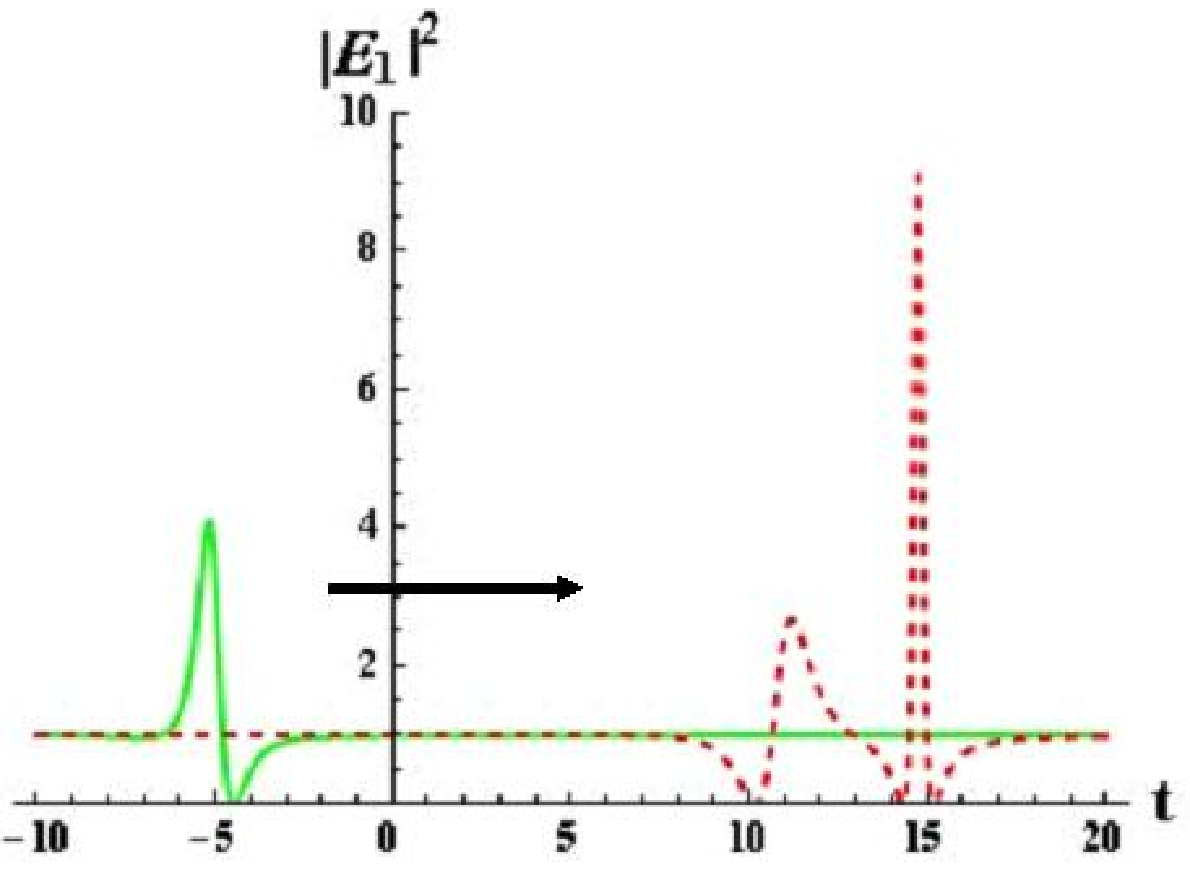}}
\hfil
\subfigure[]{\includegraphics[height=52mm,width=70mm]{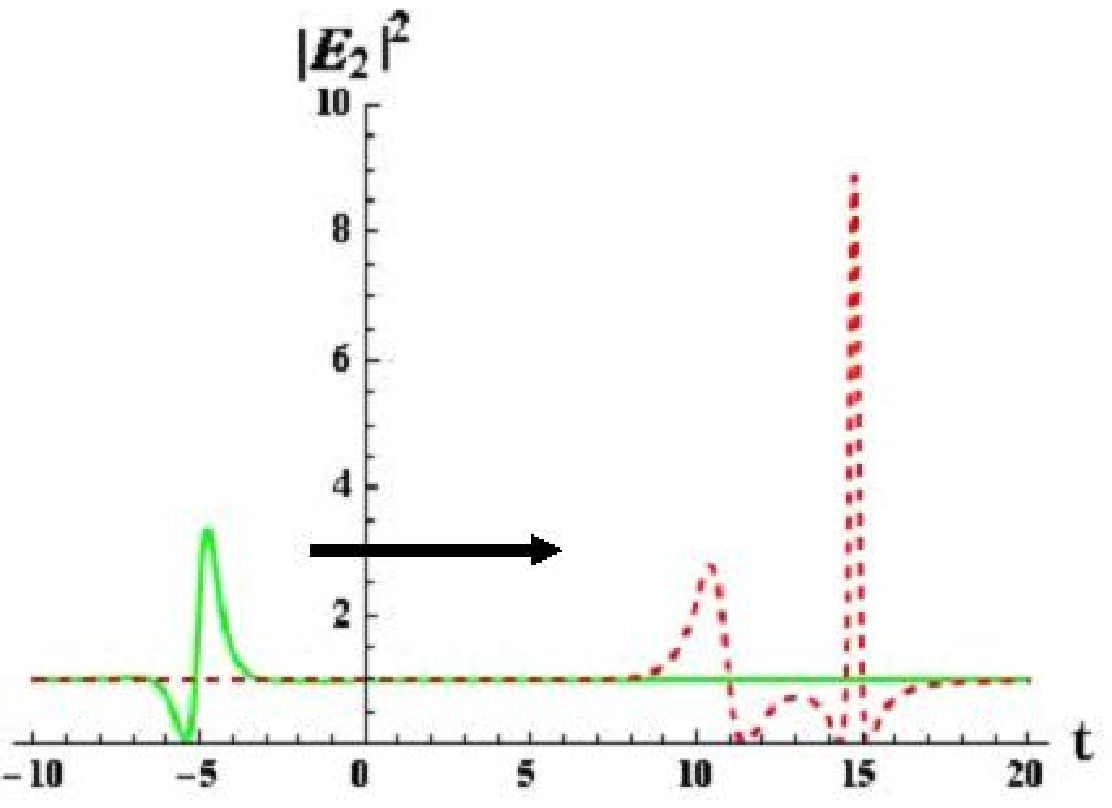}}
\caption{(color online) The dynamics of the semi-rational solution
which describe the process that one D-AD soliton pair splits into a
new D-AD one and a W-shaped soliton. (a) The evolution of localized
waves in $E_1$. (b) The evolution of localized waves in $E_2$. (c)
The distribution shape of the initial wave (solid green line) and
the subsequent waves after splitting (dashed red line) in $E _1$.
(d) The distribution shape of the initial wave (solid green line)
and the subsequent waves after splitting (dashed red line) in $E
_2$. }
\end{figure}

\emph{ Semi-rational solution}---When  $A_2\neq 0$
 and $A\neq0$, the solution corresponds to semi-rational solution, which
 describe the dynamics of a D-AD soliton pair split to a new D-AD and a localized waves with
 stable one hump with two valleys structure. With $A_1=0,A_2=A=1,c_1=c_2=1
 $, and $\epsilon=\frac{1}{12}$, the semi-rational solution can be
 simplified as
\begin{eqnarray}
E_{12}&=&\frac{4 e^{\frac{28 z}{3}}+8(1+2 t-6 z) e^{2 t+\frac{14
z}{3}} +F[t,z] e^{4 t} }{4 e^{\frac{28 z}{3}}+
[1+8 (t-3 z)^2+4(t-3 z)]e^{4 t}}\nonumber\\
&& \cdot \exp{[2i(t-\frac{2z}{3})]},\\
E_{22}&=&\frac{4 e^{\frac{28 z}{3}}-8  (1+2 t-6 z)e^{2 t+\frac{14
z}{3}} + F[t,z] e^{4 t}}{4 e^{\frac{28 z}{3}}+ [1+8 (t-3 z)^2+4(t-3
z)]e^{4
t}} \nonumber\\
&& \cdot \exp{[2i(t-\frac{2z}{3})]},
\end{eqnarray}
where $F[t,z]=1-8 (t-3 z)^2-4(t-3 z)$. Based on the solutions, we
can show the evolution of corresponding localized waves in Fig. 2.
In Fig. 2(a) and (b), we can see that the initial localized wave
($z\rightarrow -\infty $) possesses one hump and one valley
structure and the structure do not vary with the propagation
distance before the location where it splits. Therefore, it can be
seen a D-AD pair localized wave. Near a location $z=0.5$, it splits
into one new D-AD pair and a new localized waves, shown in Fig. 2
(c) and (d). The new D-AD pair is different from the initial one.
The distribution of the hump and valley is inverse to the initial
one's. The soliton's peak and width is variable too. The new
localized waves denote one anti-dark soliton with two dark soliton
around structure which is identical with the distribution structure
of NLS RW with maximum peak. We call it W-shaped soliton in this
paper since the shape is kept well with time after it emerges. This
comes from its semi-rational character.

\begin{figure}[htb]
\centering
\subfigure[]{\includegraphics[height=56mm,width=75mm]{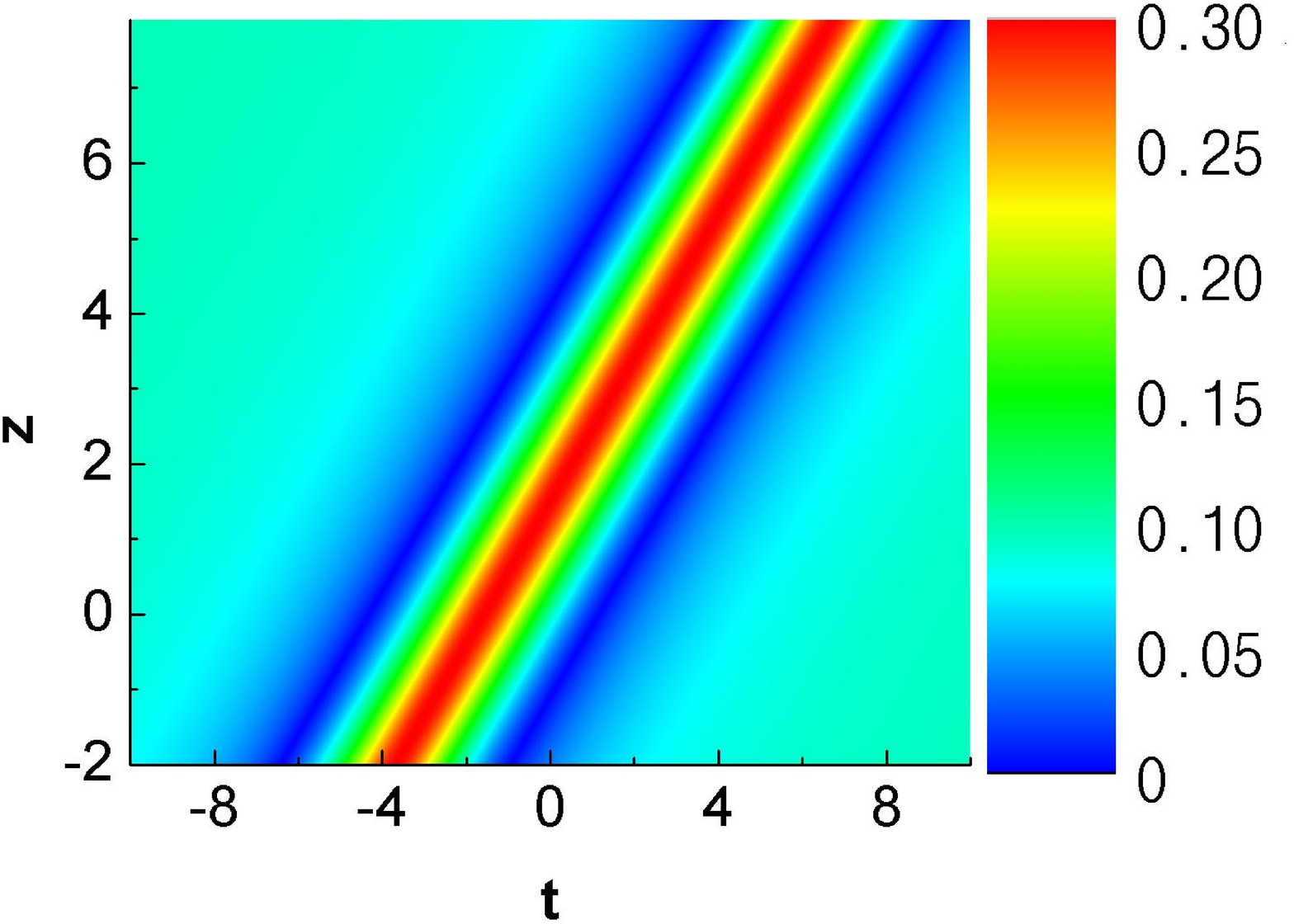}}
\hfil
\subfigure[]{\includegraphics[height=52mm,width=75mm]{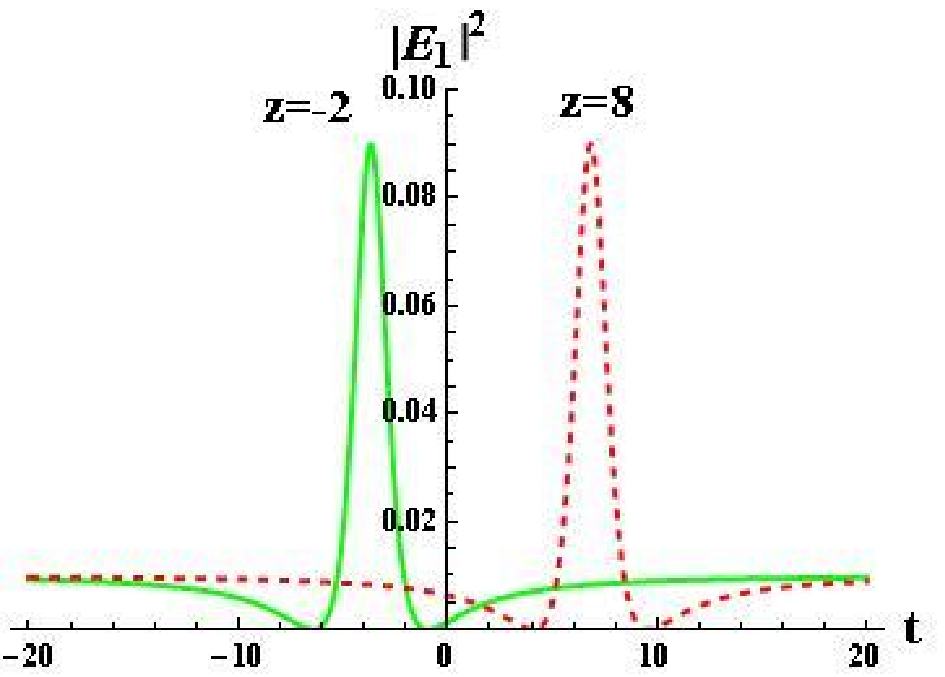}}
\caption{(color online) (a) The evolution of W-shaped
soliton in $E_1$ which is described by the rational solution. (b)
The cut plots of the soliton at two propagation distances $z=-2$
(solid green line) and $z=8$ (dashed red line). The parameter are
$A_1=0,A_2=1, A=0,c_1=0.1, c_2=0.2
 $, and $\epsilon=\frac{1}{12}$. }
\end{figure}

\emph{ Rational solution}--- When $A_2\neq0$ and $A=0$, we can
obtain rational solutions from the generalized solution. The
rational solutions describe the dynamics of W-shaped
localized waves, shown in Fig. 3. Since the density evolution of the
localized waves in two components are similar, we just show the wave
in $E_1$. The highest peak value is nonuple than the value of
background. This property is similar to the NLS RW with highest
peak. However, the whole evolution is quite distinctive from each
other. The structures are kept very well, unlike the NLS RW
disappear quickly, which is the reason why we call then RW-shaped
soliton. Moreover, the rational solutions of the two components  are
identical when $c_1=c_2$. Namely, the vector solutions can be seen
as scalar rational solution of S-S equation. With $A_1=0,A_2=1,
A=0,c_1=c_2=1
 $, and $\epsilon=\frac{1}{12}$, the rational solution can be
 simplified as
\begin{eqnarray}
E_{13}= E_{23}&=&\frac{(2t-6z-1)^2-12 (t-3z)^2}{(2t-6z+1)^2+4 (t-3z)^2}\nonumber\\
&&\cdot \exp{[2i(t-\frac{2z}{3})]},
\end{eqnarray}
Its dynamic behavior is quite distinctive from the one obtained in
\cite{rws}, which is similar to the standard NLS RW, but has two
peaks. Therefore, the rational solution obtained here is a new type
rational solution of scalar S-S equation.

\subsection{Combined W-shaped solitons and dark W-shaped solitons}
 The frequency difference has real physical effect in the two-mode fiber. Therefore, we consider the case with
 $k_1=0$ and $k_2\neq0$ for simplicity. The effects can be discussed by varying $k_2$ conveniently. With
 the
 requirements on the backgrounds and the spectral parameter as
 follows
\begin{eqnarray}
k_2&=&\sqrt{\frac{2}{5}} \sqrt{c_1^2+c_2^2},\nonumber\\
c_1&=&\frac{c_2}{2},\nonumber\\
\lambda&=&i \sqrt{\frac{2}{3}} \sqrt{2 c_1^2+2 c_2^2+k_2^2},
\end{eqnarray}

\begin{figure}[htb]
\centering
\subfigure[]{\includegraphics[height=56mm,width=75mm]{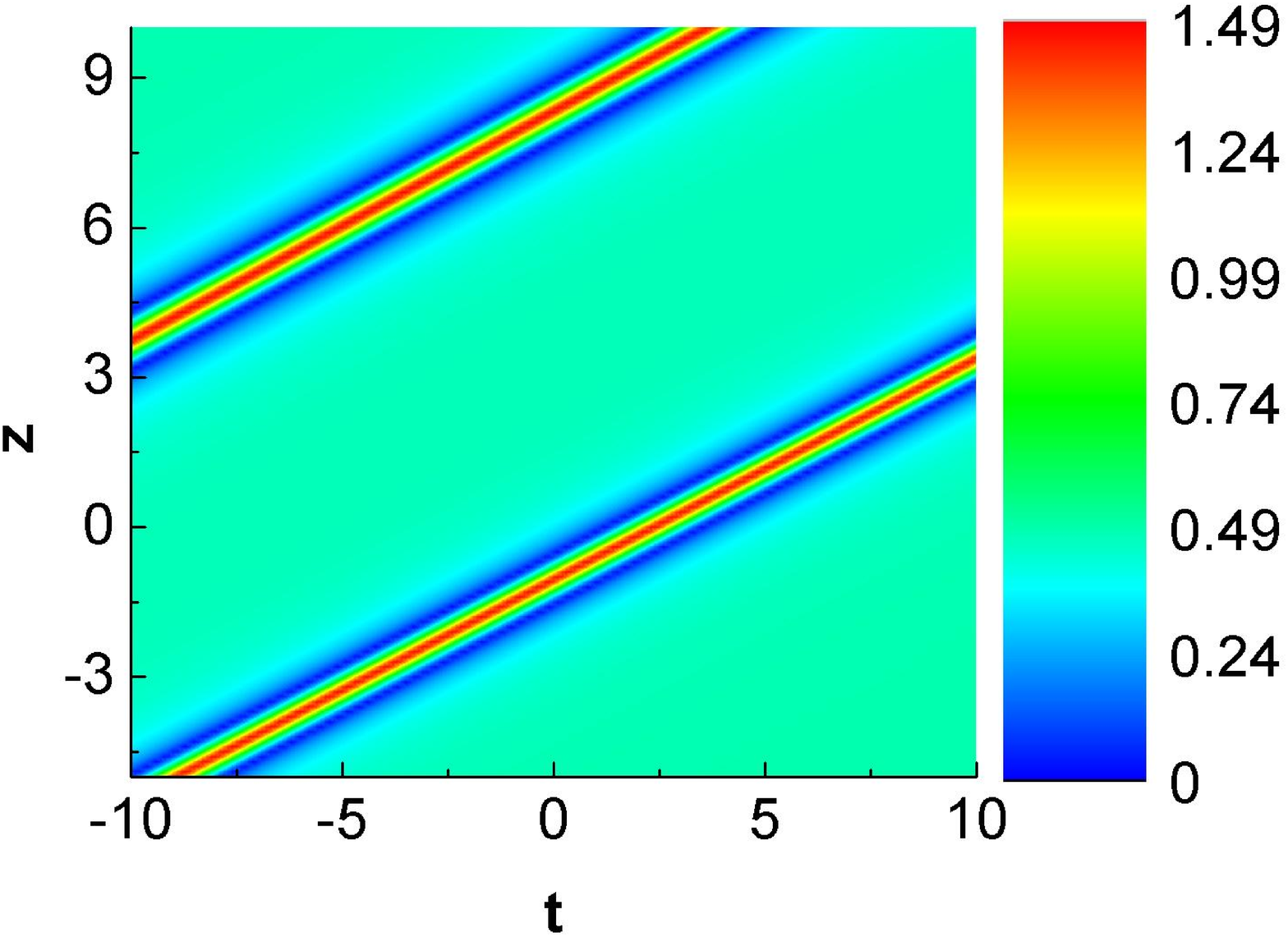}}
\hfil
\subfigure[]{\includegraphics[height=56mm,width=75mm]{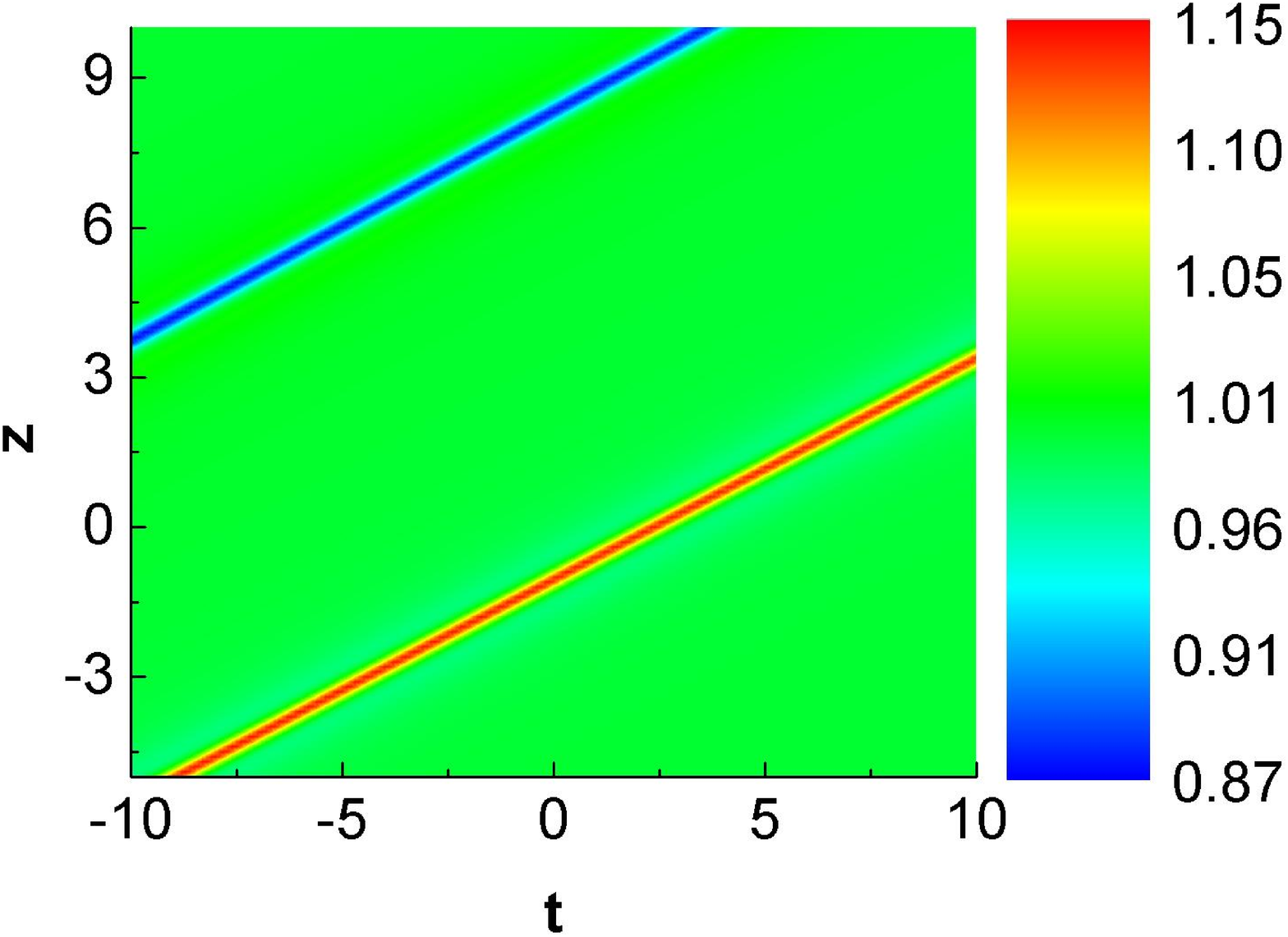}}
\hfil
\subfigure[]{\includegraphics[height=52mm,width=70mm]{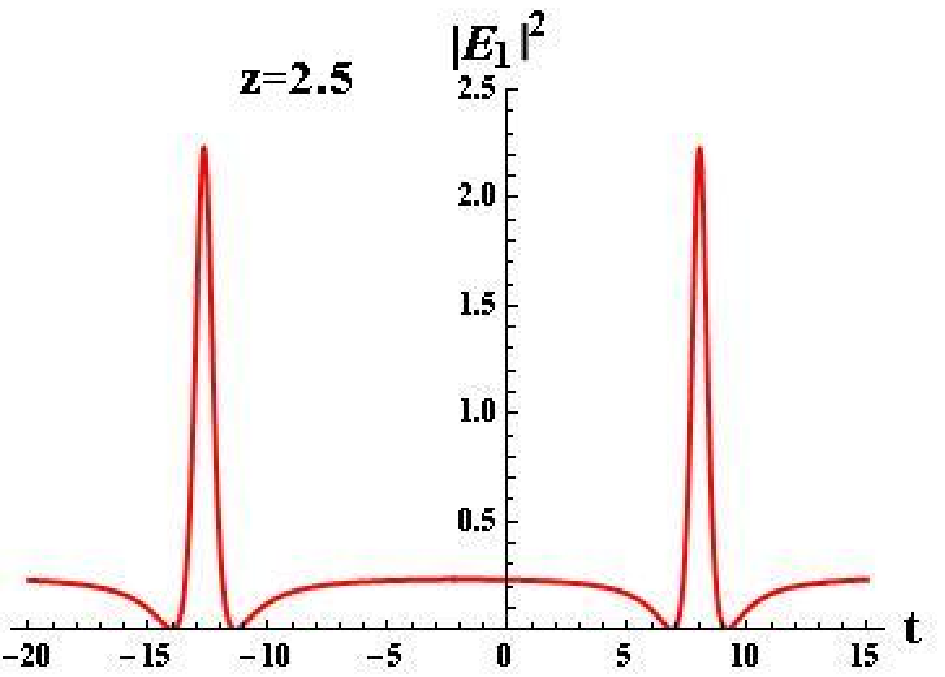}}
\hfil
\subfigure[]{\includegraphics[height=52mm,width=70mm]{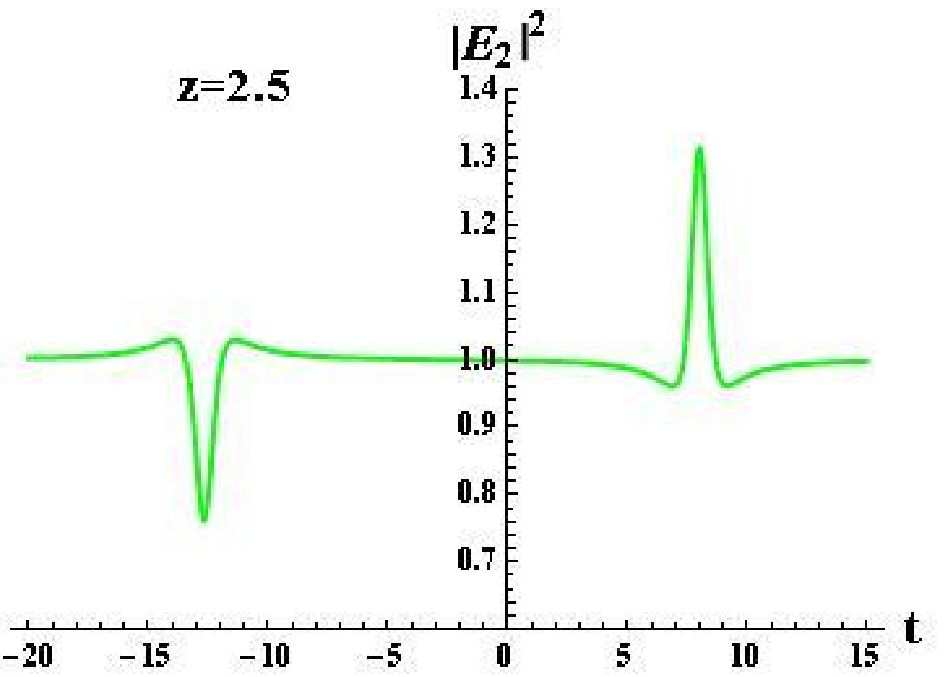}}
\caption{(color online) The dynamics of the localized wave solution
with $A_1= A_2=A_5=0$, $A_3=10$, $A_4=1$, and $\epsilon =0.2$. (a)
The evolution of localized waves in $E_1$. (b) The evolution of
localized waves in $E_2$. (c) The cut plot of the density
distribution in $E_1$. (d) The cut plot of the density distribution
in $E_2$. It is shown that there are two W-shaped solitons
in $E_1$ , there are one dark W-shaped soliton and a W-shaped soliton in $E_2$.}
\end{figure}

we obtain the generalized localized wave solution with $c_2=1$ as
\begin{eqnarray}
E_{1}&=&[\frac{1}{2}+\frac{4 \sqrt{2} \Psi_1\Psi_2^{\ast}}
{\sum_{i=1}^5|\Psi_i|^2}] \cdot
\exp{[\frac{i}{6\epsilon}(t-\frac{z}{18\epsilon})]},\nonumber\\
E_{2}&=&[1+\frac{4 \sqrt{2} \Psi_1\Psi_4^{\ast}}
{\sum_{i=1}^5|\Psi_i|^2}]\cdot
\exp{[\frac{i}{6\epsilon}(t-\frac{z}{18\epsilon})]}\nonumber\\
&&\cdot \exp{[-\frac{i (z-12 t \epsilon +84 z \epsilon ^2)}{12
\sqrt{2} \epsilon }]},
\end{eqnarray}
where
\begin{widetext}
\begin{eqnarray}
\Psi_1&=& \left(-6 \sqrt{2}A_2+12 \sqrt{2} A_4+12 A_5 +12 A_4 T+12
\sqrt{2} A_5 T-3 \sqrt{2} A_4 T^2+6 A_5 T^2
-\sqrt{2} A_5 T^3\right. \nonumber\\
&&\left.-90 A_4 z \epsilon -120 \sqrt{2} A_5 z \epsilon +54 \sqrt{2}
A_4 T z \epsilon -90 A_5 T z \epsilon +27 \sqrt{2}A_5 T^2 z \epsilon
-243
\sqrt{2} A_4 z^2 \epsilon ^2\right. \nonumber\\
&&\left.+324 A_5 z^2 \epsilon ^2-243 \sqrt{2} A_5 T z^2 \epsilon ^2
+729 \sqrt{2} A_5 z^3 \epsilon ^3-6 A_3
(-2+\sqrt{2} T-9 \sqrt{2} z \epsilon)\right)\cdot \frac{1}{6} e^{-\frac{5 T+8 z \epsilon }{5 \sqrt{2}}}, \nonumber\\
\Psi_2&=&  \left(-3 \sqrt{2} A_5 +i A_1 e^{4 \sqrt{2} z
\epsilon-\frac{T}{\sqrt{2}} }+\frac{1}{6}  [6 A_2+3 A_4 T^2 -9
\sqrt{2} A_4 z \epsilon -6 A_5 z \epsilon -54 A_4 T z
\epsilon -9 \sqrt{2} A_5 T  z \epsilon \right. \nonumber\\
&&\left.-27 A_5 T^2 z \epsilon +243 A_4 z^2 \epsilon ^2+81 \sqrt{2}
A_5 z^2 \epsilon ^2+A_5 T^3 +243 A_5 T z^2 \epsilon ^2-729 A_5 z^3 \epsilon ^3+6 A_3 (T-9 z \epsilon )]\right. \nonumber\\
&&\left. +2 (A_4+A_5 T-9 A_5 z \epsilon )-2 \sqrt{2} A_3-2 \sqrt{2}
A_4 T-\sqrt{2} A_5 T^2+18 \sqrt{2} A_4 z \epsilon +6 A_5 z \epsilon \right. \nonumber\\
&&\left.+18 \sqrt{2} A_5 T z \epsilon -81 \sqrt{2} A_5 z^2 \epsilon
^2\right) e^{-\frac{5 T+8 z \epsilon}{5\sqrt{2}}}, \nonumber\\
\Psi_4&=& \left((6+6 i) A_2-(18-6 i) A_4+6 \sqrt{2} A_5 -(6+12 i)
\sqrt{2} A_4 T +(174-24 i) A_5 z \epsilon
\right. \nonumber\\
&&\left.-(18-6 i) A_5 T+(3+3 i) A_4 T^2 -(3+6 i) \sqrt{2} A_5
T^2+(1+i) A_5 T^3+(45+99 i) \sqrt{2} A_4 z \epsilon \right. \nonumber\\
&&\left.-(54+54 i) A_4 T z \epsilon +(45+99 i) \sqrt{2} A_5 T z
\epsilon -(27+27 i) A_5 T^2 z \epsilon +(243+243 i) A_4 z^2 \epsilon
^2\right. \nonumber\\
&&\left.-(162+405 i) \sqrt{2} A_5 z^2 \epsilon ^2 +(243+243 i)A_5 T
z^2 \epsilon ^2
-6 A_3 [(1+2 i) \sqrt{2}-(1+i) T+(9+9 i) z \epsilon ]\right. \nonumber\\
&&\left.-(729+729 i)A_5 z^3 \epsilon ^3 \right ) \frac{1}{6}
e^{-\frac{5 T+8 z \epsilon }{5 \sqrt{2}}},\nonumber
\end{eqnarray}
\end{widetext}
$\Psi_3=\Psi_2^{*}$ and $\Psi_5=\Psi_4^{*}$, where
$T=t-\frac{z}{12\epsilon}$, and $A_j (j=1,2,3,4,5)$ are real
numbers. The constrains condition is satisfied. When $A_1\neq0$
$A_2\neq0$ and $A_{3,4,5}=0$, the solution become a vector antidark
-dark soliton, for which there is an antidark soliton in $E_1$ and a
dark soliton in $E_2$. When $A_1=0$, we can get rational solution
which can be used to describe the combined W-shaped solitons. When
$A_5=0$, $A_4=0$ and $A_3\neq0$, we can get a vector localized waves
for which there is a W-shaped soliton in $E_1$, and a plane wave
with no localized wave in $E_2$. The W-shaped soliton in $E_1$ is
similar to the one in Fig. 3.

\begin{figure}[htb]
\centering
\subfigure[]{\includegraphics[height=56mm,width=75mm]{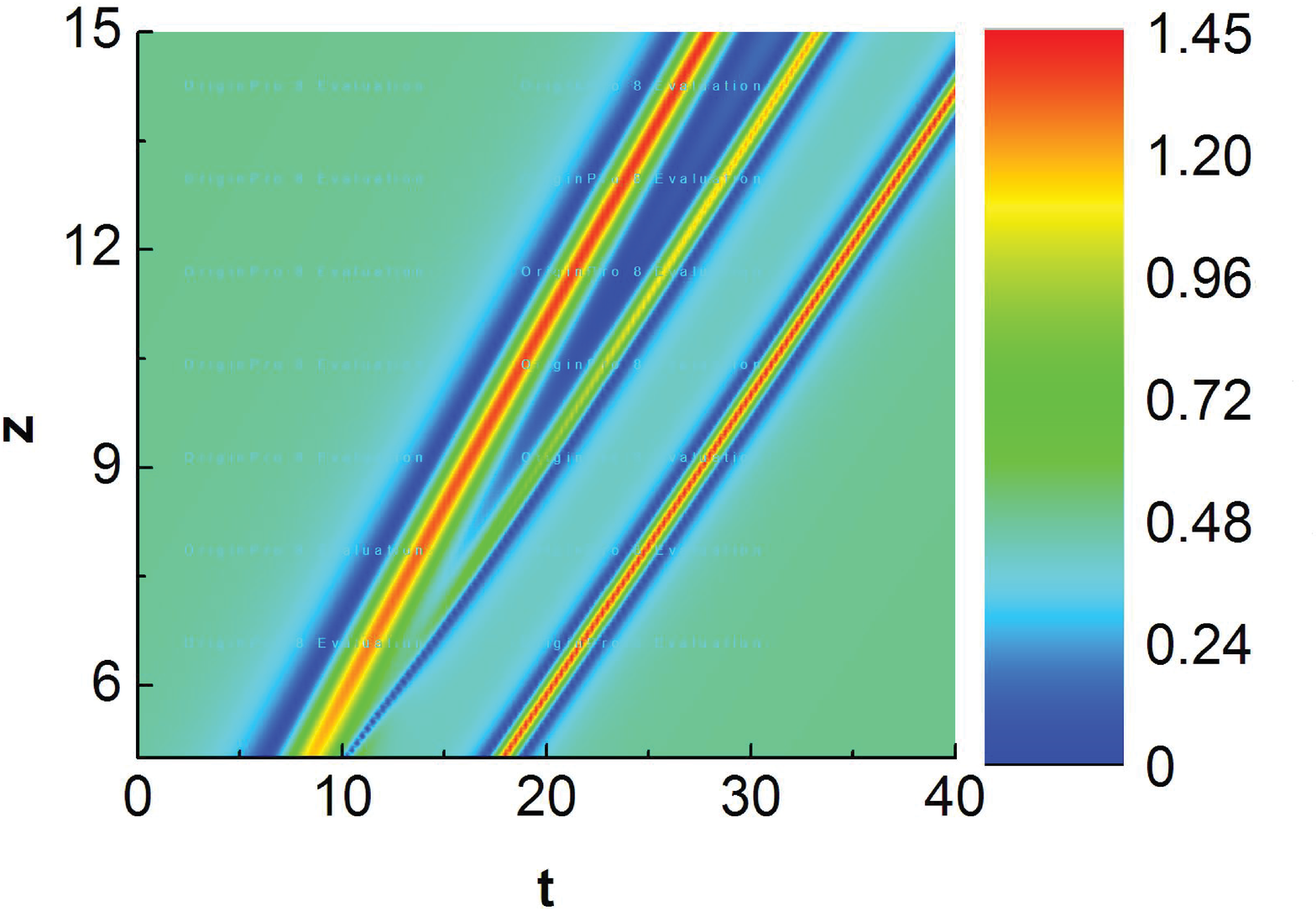}}
\hfil
\subfigure[]{\includegraphics[height=56mm,width=75mm]{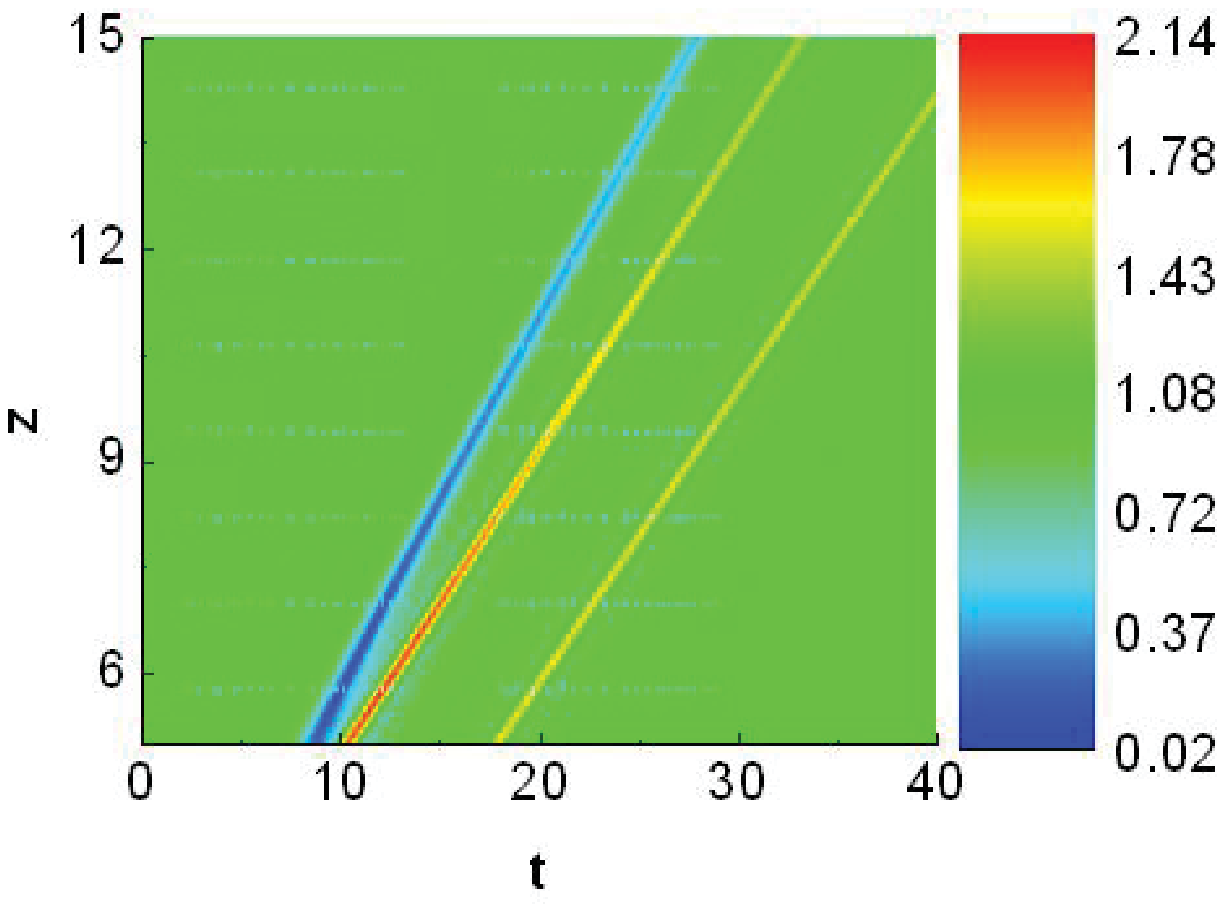}}
\hfil
\subfigure[]{\includegraphics[height=52mm,width=70mm]{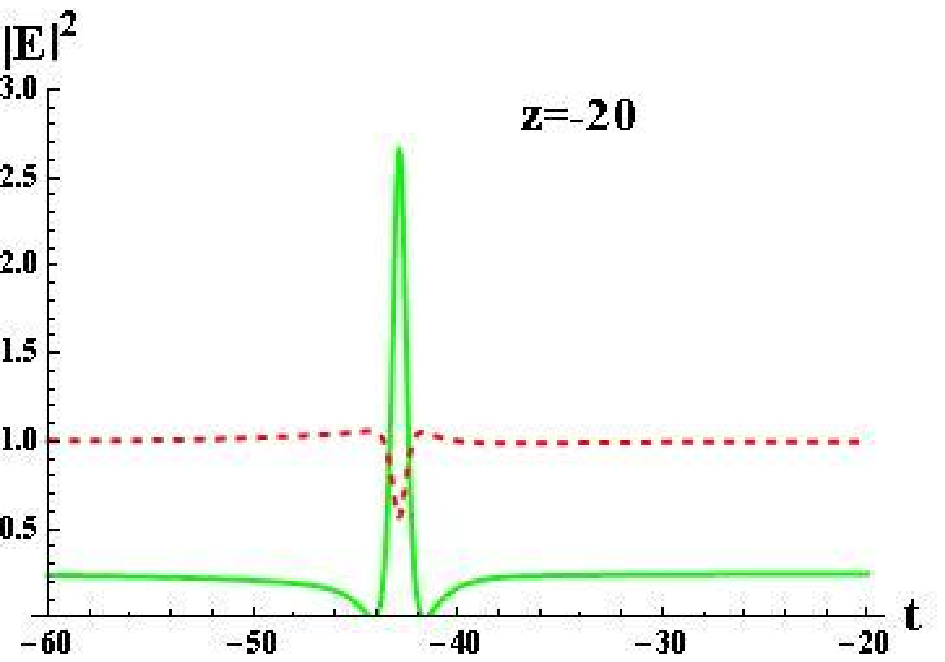}}
\hfil
\subfigure[]{\includegraphics[height=52mm,width=70mm]{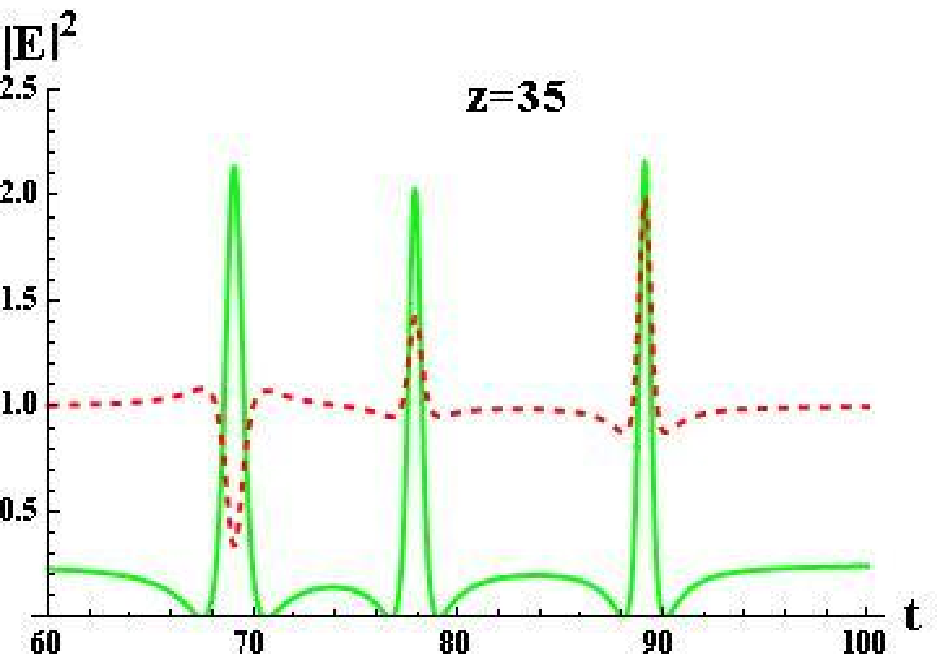}}
\caption{(color online) The dynamics of the localized wave solution
with $A_1= A_2=0$, $A_3=A_4=1$, $A_5=1$, and $\epsilon =0.2$. (a)
The evolution of localized waves in $E_1$. (b) The evolution of
localized waves in $E_2$. (c) The cut plot at $z=-20$ of the density
distribution in $E_1$(solid green line) and $E_2$ (dashed red line).
 (d) The cut plot
at $z=35$ of the density distribution in $E_1$(solid green line) and
$E_2$ (dashed red line).  }
\end{figure}

When $A_4\neq 0$ and $A_5=0$, there is a combined localized wave in
$E_1$ which is consist of two W-shaped soliton(shown in Fig. 4 (a)
and (c)), and a combined localized wave in $E_2$ which is consist of
a W-shaped soliton and a dark W-shaped soliton (shown in Fig. 4(b)
and (d)). The dark W-shaped soliton has a similar shape with the
dark RW with smallest value in coupled NLS \cite{Zhao2}. Moreover,
the dark W-shaped distribution can be kept well with evolution.
Therefore, we call it dark W-shaped soliton. When $A_5\neq0$, there
is a combined localized wave in $E_1$ which is consist of three W
-shaped solitons. The initial localized wave is one W-shaped
soliton (green solid line in Fig. 5(c)), it splits into three
W-shaped solitons, and the three soliton's shapes are a bit
different from the standard RW with the maximum peak, see Fig. 5 (a)
and green solid line in (d).  The initial localized wave is a dark
W-shaped soliton(red dashed line in Fig. 5 (c)) in $E_2$, and it
splits into the one dark and two W-shaped soliton, see Fig. 5(b)
and red dashed line in (d). The interaction between them can be
observed through varying the parameters in the generalized solution.

\section{discussion and conclusion}
In this paper, we find there are many new types of localized waves
existed in the two-mode nonlinear fiber with these usual high-order
effects, such as D-AD pair, W-shaped soliton, dark W-shaped
soliton, and the combined waves of them. For coupled S-S model, some
rational solutions do not necessarily correspond to RW behavior
which seems to appear from nowhere and disappear without any trace.
Their behavior are quite distinct from the ones in coupled NLS
equations \cite{Baronio,Ling2,Zhao2,Zhao3}. It is well known that
the localized waves on nonzero background possesses breathing
dynamics in coupled NLS equations. But the localized waves here do
not breath anymore with these high-order effects. This means that
the high-order effects can be used to compress the breathing
behavior.

Considering the results in \cite{SS,rws, K. Nakkeeran, D.
Mihalache}, we can know that the solutions obtained here are in the
branches which are different from them. It is still need to find out
one whole picture for all branches. We just present localized wave
solutions which can be written exactly and explicitly. In fact,
there are many other types generalized localized wave solutions for
the coupled model, since the matrix in Lax-pair is $5\times5$. We
will further study on this direction.

\section*{Acknowledgments}
This work is supported by the National Fundamental Research Program
of China (Contact 2011CB921503), the National Science Foundation of
China (Contact Nos. 11274051, 91021021).

\end{document}